\documentclass[12pt]{article}

\usepackage{amsfonts}
\usepackage{amssymb}
\usepackage{amsmath}
\usepackage{dsfont}
\usepackage{hyperref}
\usepackage{slashed}
\usepackage{empheq}
\usepackage{multirow}
\usepackage{fancyhdr}
\usepackage{subfigure}

\usepackage{amsmath,amssymb,amsfonts,amsthm,dcolumn}
\usepackage[margin=-5pt]{caption}
\usepackage{graphicx,wrapfig}
\usepackage{tensor}

\def \ov {\over}

 \def\ep {\epsilon}

\def \ep {\epsilon}
\def \te {\tilde \epsilon}

\newcounter{subequation}[equation]

\newcommand{\be}{\begin{equation}}
\newcommand{\ee}{\end{equation}}
\newcommand{\eel}[1]{\label{#1}\end{equation}}
\newcommand{\bea}{\begin{eqnarray}}
\newcommand{\eea}{\end{eqnarray}}
\newcommand{\eeal}[1]{\label{#1}\end{eqnarray}}

\makeatletter

\def\thesubequation{\theequation\@alph\c@subequation}
\def\@subeqnnum{{\rm (\thesubequation)}}
\def\slabel#1{\@bsphack\if@filesw {\let\thepage\relax
   \xdef\@gtempa{\write\@auxout{\string
      \newlabel{#1}{{\thesubequation}{\thepage}}}}}\@gtempa
   \if@nobreak \ifvmode\nobreak\fi\fi\fi\@esphack}
\def\subeqnarray{\stepcounter{equation}
\let\@currentlabel=\theequation\global\c@subequation\@ne
\global\@eqnswtrue \global\@eqcnt\z@\tabskip\@centering\let\\=\@subeqncr

$$\halign to \displaywidth\bgroup\@eqnsel\hskip\@centering
  $\displaystyle\tabskip\z@{##}$&\global\@eqcnt\@ne
  \hskip 2\arraycolsep \hfil${##}$\hfil
  &\global\@eqcnt\tw@ \hskip 2\arraycolsep
  $\displaystyle\tabskip\z@{##}$\hfil
   \tabskip\@centering&\llap{##}\tabskip\z@\cr}
\def\endsubeqnarray{\@@subeqncr\egroup
                     $$\global\@ignoretrue}
\def\@subeqncr{{\ifnum0=`}\fi\@ifstar{\global\@eqpen\@M
    \@ysubeqncr}{\global\@eqpen\interdisplaylinepenalty \@ysubeqncr}}
\def\@ysubeqncr{\@ifnextchar [{\@xsubeqncr}{\@xsubeqncr[\z@]}}
\def\@xsubeqncr[#1]{\ifnum0=`{\fi}\@@subeqncr
   \noalign{\penalty\@eqpen\vskip\jot\vskip #1\relax}}
\def\@@subeqncr{\let\@tempa\relax
    \ifcase\@eqcnt \def\@tempa{& & &}\or \def\@tempa{& &}
      \else \def\@tempa{&}\fi
     \@tempa \if@eqnsw\@subeqnnum\refstepcounter{subequation}\fi
     \global\@eqnswtrue\global\@eqcnt\z@\cr}
\let\@ssubeqncr=\@subeqncr
\@namedef{subeqnarray*}{\def\@subeqncr{\nonumber\@ssubeqncr}\subeqnarray}

\@namedef{endsubeqnarray*}{\global\advance\c@equation\m@ne
                           \nonumber\endsubeqnarray}

\makeatletter \@addtoreset{equation}{section} \makeatother
\renewcommand{\theequation}{\thesection.\arabic{equation}}

\catcode`\@=11

\newcount\hour
\newcount\minute
\newtoks\amorpm \hour=\time\divide\hour by 60\minute
=\time{\multiply\hour by 60 \global\advance\minute by-\hour}
\edef\standardtime{{\ifnum\hour<12 \global\amorpm={am}
        \else\global\amorpm={pm}\advance\hour by-12 \fi
        \ifnum\hour=0 \hour=12 \fi
        \number\hour:\ifnum\minute<10
        0\fi\number\minute\the\amorpm}}
\edef\militarytime{\number\hour:\ifnum\minute<10 0\fi\number\minute}

\def\draftlabel#1{{\@bsphack\if@filesw {\let\thepage\relax
   \xdef\@gtempa{\write\@auxout{\string
      \newlabel{#1}{{\@currentlabel}{\thepage}}}}}\@gtempa
   \if@nobreak \ifvmode\nobreak\fi\fi\fi\@esphack}
        \gdef\@eqnlabel{#1}}
\def\@eqnlabel{}
\def\@vacuum{}
\def\marginnote#1{}
\def\draftmarginnote#1{\marginpar{\raggedright\scriptsize\tt#1}}
\def\draft{
        \pagestyle{plain}
        \overfullrule=2pt
        \oddsidemargin -.5truein
        \def\@oddhead{\sl \phantom{\today\quad\militarytime} \hfil
        \smash{\Large\sl DRAFT} \hfil \today\quad\militarytime}
        \let\@evenhead\@oddhead
        \let\label=\draftlabel
        \let\marginnote=\draftmarginnote
        \def\ps@empty{\let\@mkboth\@gobbletwo
        \def\@oddfoot{\hfil \smash{\Large\sl DRAFT} \hfil}
        \let\@evenfoot\@oddhead}

\def\@eqnnum{(\theequation)\rlap{\kern\marginparsep\tt\@eqnlabel}
        \global\let\@eqnlabel\@vacuum}  }

\renewcommand{\theequation}{\thesection.\arabic{equation}}
\renewcommand{\thefootnote}{\fnsymbol{footnote}}

\def\appendix#1{
  \addtocounter{section}{-3}
  \setcounter{equation}{0}
  \renewcommand{\thesection}{\Alph{section}}
  \section*{Appendix \thesection\protect\indent \parbox[t]{11.15cm}
  {#1} }
  \addcontentsline{toc}{section}{Appendix \thesection\ \ \ #1}
  }
\def \ov {\over}

 \def\ep {\epsilon}

\def \ep {\epsilon}

\def\O{\Omega}

\def\pd{\partial}

\def\a{\alpha}

\def\s{\sigma}

\def\te{\theta}

\def\ep{\epsilon}

\def\be{\begin{equation}}
\def\ee{\end{equation}}

\def \k {\kappa}

\def\te{\theta}

\newcommand{\bal}{\begin{align}}
\newcommand{\ean}{\end{align}}

\date{}
\begin{document}

\begin{titlepage}

\hfill MCTP-11-12\\

\begin{center}

{\Large \bf Chaos Rules out Integrability of Strings on $AdS_5\times T^{1,1}$}

\vskip .7 cm

\vskip 1 cm

{\large   Pallab Basu${}^1$ and Leopoldo A. Pando Zayas${}^2$}

\end{center}

\vskip .4cm
\centerline{\it ${}^1$ Department of Physics and Astronomy, University of Kentucky}
\centerline{\it  Lexington, KY 40506, USA}

\vskip .4cm \centerline{\it ${}^2$ Michigan Center for Theoretical
Physics}
\centerline{ \it Randall Laboratory of Physics, The University of
Michigan}
\centerline{\it Ann Arbor, MI 48109-1120}

\vskip .4cm
\centerline{ \it }
\centerline{\it  }

\vskip 1 cm

\vskip 1.5 cm

\begin{abstract}
We show that certain classical string configurations in $AdS_5\times T^{1,1}$ are chaotic. This answers the question of integrability of string on such backgrounds in the negative. We consider a string localized in the center of $AdS_5$ that winds around two circles of $T^{1,1}$. The corresponding dynamical system is equivalent to two coupled gravitational pendula and allows a very intuitive understanding. We find conclusive evidence of chaotic behavior by systematically analyzing the workings of the KAM theorem. We also show that the largest Lyapunov exponent is positive. 
\end{abstract}

\end{titlepage}
\setcounter{page}{1} \renewcommand{\thefootnote}{\arabic{footnote}}
\setcounter{footnote}{0}

\def \N{{\cal N}}
\def \ov {\over}

\section{ Introduction}
One of the most interesting paradigms arising from the AdS/CFT \cite{Maldacena:1997re,Witten:1998qj,Gubser:1998bc, Aharony:1999ti} correspondence has been the role of integrability (for an extensive review see \cite{Beisert:2010jr}). Integrability has modified our understanding of field theory to the point of reformulating, in some cases, the solution of a conformal field theory in terms of S-matrix elements of integrable systems. In practical terms, integrability has allowed exploration of sectors of the theory that would have been otherwise inaccessible with standard methods.

In the context of the duality between strings on $AdS_5\times S^5$ with Ramond-Ramond fluxes and ${\cal N}=4$ supersymmetric Yang-Mills (SYM) with gauge group $SU(N)$, integrability has two faces or two ways of appearing; they correspond to the two extreme values of the 't Hooft coupling. On the string theory side which admits a classical description at large 't Hooft coupling $\lambda$, integrability of the classical sigma model on $AdS_5\times S^5$ was suggested in \cite{Mandal:2002fs} and fully established with the inclusion of fermions in \cite{Bena:2003wd}. The field theoretic face of integrability originated out of the BMN limit, through the spin chain and relies on the form of the dilatation operator in an expansion around $\lambda \to 0$. The key components of the integrability story are by now nicely summarized in \cite{Beisert:2010jr}.

More generally, the gauge/gravity correspondence provides a new method of attacking field theories at strong coupling. A natural hope is that the methods used in the standard duality could be developed to understand more realistic field theories and eventually apply them to theories in the same universality class of QCD. In this direction a first step was taken by Klebanov and Witten who considered a model with reduced supersymmetry. By taking the Maldacena limit of a stack of D3 branes at the tip of a conifold they arrived at a supergravity background of the form $AdS_5\times T^{1,1}$ \cite{Klebanov:1998hh}. This background is dual to an ${\cal N}=1$ superconformal field theory.

The BMN limit of $AdS_5\times T^{1,1}$ was a clear indication that some of the properties of ${\cal N}=4$ supersymmetric Yang-Mills could be found in certain sectors of superconformal ${\cal N}=1$ field theories with AdS dual \cite{PandoZayas:2002rx,Itzhaki:2002kh,Gomis:2002km}. More generally, it was shown in \cite{Itzhaki:2002kh} that the corresponding BMN sector can be found in any gravity dual of the form $AdS_5\times X^5$ where $X^5$ is a Sasaki-Einstein $5-$manifold. Many classical string configurations found in $AdS_5\times S^5$  have also been extended to analogous configurations in $AdS_5\times T^{1,1}$. For example, one can construct classical configurations of spinning strings in $T^{1,1}$ \cite{Kim:2003vn,Wang:2005baa}. However, many unsuccessful attempts have been made to demonstrate that the classical string in $AdS_5\times T^{1,1}$ is integrable. In particular it can be established that the construction of \cite{Bena:2003wd} does not extend to $AdS_5\times T^{1,1}$ due to the fact that $T^{1,1}$ is not a maximally symmetric coset space. The common consensus has been that some integrable structure is present but full integrability is, at least, elusive.

Let us also remark that attacking the problem from the field theory side is rather hard as the corresponding field theory is defined  at a strongly coupled conformal fixed point. This situation prevents a perturbative analysis from taking place. In other words, the fact that the fields have large anomalous dimensions in the conformal fixed point prevents a perturbative computation. An interesting study performed in \cite{Berenstein:2004ys} showed that the requirement of integrability of the full one loop dilatation operator in the scalar sector places very strong constraints on the field theory, so that the only soluble models correspond essentially to orbifolds of ${\cal N }=4$.

In this paper we show that in general the classical string in $AdS_5\times T^{1,1}$ can not be integrable. We settle this question not by our failure to show that the theory is integrable as has been the standard path so far. Rather, we present a simple classical configuration displaying chaotic behavior. The configuration in question is a string that wounds along two of the directions of $T^{1,1}$; this configuration is inspired by the ring string in the Schwarzschild black hole in $AdS_5$ considered in \cite{Zayas:2010fs} but it is a lot simpler. As a result of the choice of Ansatz, the motion of the string reduces to the motion of two coupled oscillators that we analyze in detail. In a forthcoming work a similar albeit slightly more complicated construction is used to show non-integrability of string dynamics in AdS soliton background \cite{Basu:2011aa}.

The structure of the paper is as follows. In section \ref{sec:analysis} we present the Ansatz, its corresponding equations of motion and proceed to analyze various swaths of the phase space showing that the system presents some islands of integrability but is generically chaotic. To prove that the motion is chaotic we focus on the form of some Poincar\`e sections; we use them to exemplify the workings of the Kolmogorov-Arnold-Moser (KAM) theorem. Namely, we show explicitly how the KAM tori get increasingly scattered as we change the energy. We also present a study of the the largest Lyapunov exponent in section \ref{sec:analysis}. Section \ref{sec:fieldtheory} contains a few comments about the field theory dual and section \ref{sec:conclusions} contains our conclusions.

\section {Wrapped classical  strings in $ T^{1,1}$}\label{sec:analysis}
We start by considering the $AdS_5\times T^{1,1}$ background with metric
\bea
ds^2&=& R^2\left(-\cosh^2\rho\,\, dt^2 + d\rho^2 + \sinh^2\rho\, d\O_3^2
\right.  \nonumber \\
&+& \left. \frac16\sum\limits_{i=1}^2( d\te_i^2 +\sin^2\te_i d\phi_i^2)+
  \frac19(d\psi+\sum\limits_{i=1}^2\cos\te_id\phi_i)^2\right).
\eea
The Polyakov action for the string in the conformal gauge
\be
S={1\over 4\pi\a'}\int d^2\s\, G_{ij} \pd_a X^i \pd^a X^j,
\ee
must be supplemented by the constraints
\bea
G_{ij}\bigg[\pd_\tau X^i\, \pd_\tau X^j+\pd_\s X^i\, \pd_\s X^j \bigg]=0, \qquad
G_{ij}\pd_\tau X^i\, \pd_\s X^j=0.
\eea

We will consider solutions of the string localized in $AdS_5$ at $\rho=0$. The string will wrap the directions $\phi_1$ and $\phi_2$ in
$T^{1,1}$, a natural Ansatz will be
\bea
\phi_1&=&\alpha_1\sigma, \quad \phi_2=\alpha_2\sigma, \nonumber \\
t&=&t(\tau), \quad \psi=\psi(\tau), \quad \theta_i=\theta_i(\tau).
\eea

Some of the coordinates involved have simple solutions $\dot{t}=E$ and $\dot{\psi}=J$, the nontrivial equations are:

\bea
\ddot{\theta}_1+\alpha_1 \, \sin\theta_1 \left(\frac{\alpha_1}{3}\cos\theta_1 -\frac{2\alpha_2}{3}\cos\theta_2\right)&=&0, \nonumber \\
\ddot{\theta}_2+\alpha_2 \, \sin\theta_2\left(\frac{\alpha_2}{3}\cos\theta_2 -\frac{2\alpha_1}{3}\cos\theta_1\right)&=&0, \nonumber \\
\eea

The constraints yield

\be
E^2=\frac{1}{9} J^2 + \frac{1}{9}\sum\limits_{i=1}^2\alpha_i^2 + \frac{1}{6}\sum\limits_{i=1}^2\dot{\theta}_i^2 +\frac{1}{18}\sum\limits_{i=1}^2\alpha_i^2 \sin^2\theta_i +\frac{2}{9}\alpha_1\alpha_2\cos\theta_1\cos\theta_2.
\ee

It is easy to identify a natural classical analogy for our system. The above expression matches two gravitational pendula coupled through an interaction of the form $\cos\theta_1 \, \cos\theta_2$.  We will exploit this mechanical analogy throughout our analysis.

One of the most important results of XIX century mechanics was the identification of the role of the {\it full} phase space for characterizing a system, rather than the role of single trajectories. This is precisely the route that we will follow in the coming sections. The case for studying the full phase space has been made previously in the context of classical solution in the AdS/CFT correspondence \cite{PandoZayas:2008fw}.

\subsection{Dynamics of the system}
In this section we discuss general aspects of the dynamics of the system.

If one of the $\alpha_i$ vanishes then the system reduces to a one dimensional model and hence is integrable in terms of Jacobi elliptic functions. Namely, for  $\alpha_2=0$ we have  $\theta_2= a \tau +b$ and
\be
\ddot{\theta}_1+\frac{\alpha^2_1}{6} \sin 2\theta_1=0 \longrightarrow \theta_1(\tau)={\rm am}((\tau+C_2)\sqrt{\alpha^2_1/6+C_1}, \frac{\alpha^2_1/3}{\alpha^2_1/6+C_1}).
\ee
The motion also becomes one dimensional if we start from $sin(\theta_1)=0,\dot \theta_1=0$. Same holds for $\theta_2$.

To investigate a more generic situation we look at the  relevant potential term

\be\label{eq:potential}
V(\theta_1,\theta_2)=\frac{1}{18} \alpha_1^2 \sin ^2\left(\theta _1\right)+\frac{2}{9}\alpha_1\,\alpha_2 \cos \left(\theta _1\right) \cos \left(\theta_2\right)+\frac{1}{18} \alpha_2^2 \sin ^2\left(\theta _2\right).
\ee

It can be easily seen from the eoms that only the ration the ratio $\frac{\alpha_1}{\alpha_2}$ matters and both may be chosen to be positive. Local extrema for the above potential exist for ($\alpha_1,\alpha_2>0$),

\begin{align}
\sin(\theta_1)&=0,\sin(\theta_2)=0 \\
\cos(\theta_1)&=0,\cos(\theta_2)=0 \\
\sin(\theta_1)&=0, \cos(\theta_2)=2 \frac{\alpha_2}{\alpha_1}, \text{ when } |2 \frac{\alpha_2}{\alpha_1}<1| \\
\cos(\theta_1)&= 2 \frac{\alpha_1}{\alpha_2},\sin(\theta_1)=0, \text{ when } |2 \frac{\alpha_1}{\alpha_2}<1|
\end{align}

\begin{figure}
\begin{center}
\includegraphics[scale=0.6]{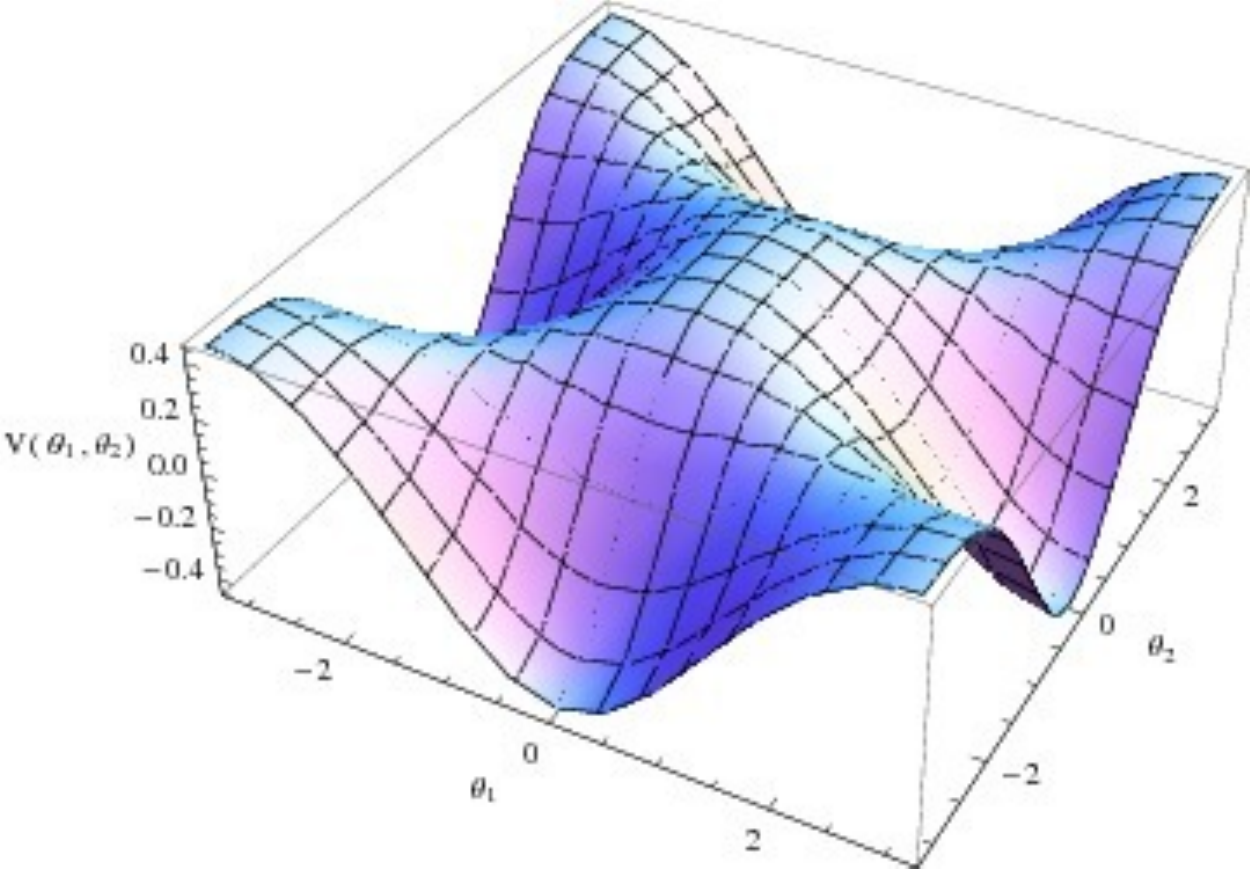}
\caption{Plot of the potential $V(\theta_1,\theta_2)$ with $\alpha_1=1$ and $\alpha_2=2$.}\label{fig:vplot}
\end{center}
\end{figure}

\begin{figure}
\begin{center}
\includegraphics[scale=0.6]{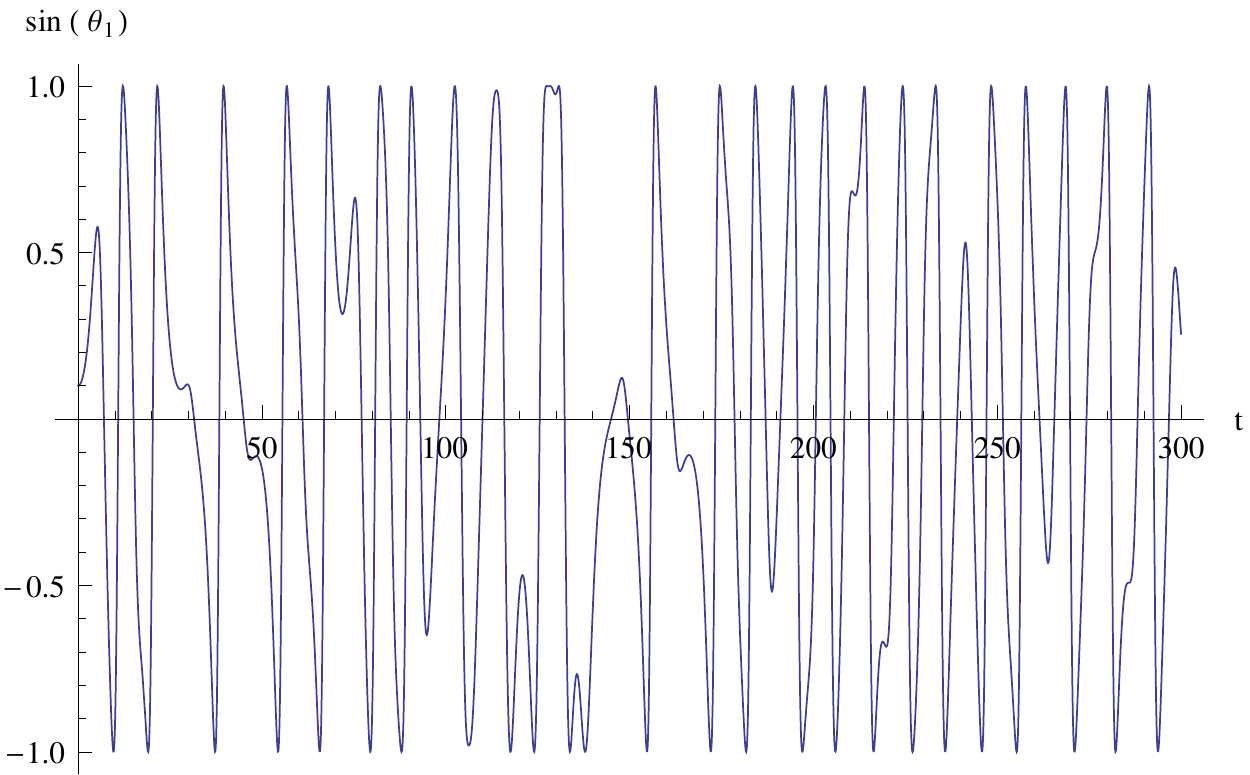}
\caption{Plot of $\sin(\theta_1(t))$ with $\alpha_1=1$ and $\alpha_2=2$ and a resting initial condition with $\theta_1(0)=0.2,\theta_2(0)=0.1$. The time evolution shows chaotic motion.}\label{fig:chaos}
\end{center}
\end{figure}

Among these extrema the point $\theta_1=0,\theta_2=\pi$ is the local minimum (see Fig. \ref{fig:vplot}). This suggests the existence of an stable fixed point near $\theta_1=0, \theta_2=\pi$. Near
this point fluctuations of $\theta_1$, $\theta_2$ couple in quartic order and for sufficiently small value of those fluctuations the system behaves like a pair of decoupled harmonic oscillators. For such a motion we observe a nice periodic behaviour. For higher values of the energy the coupling between oscillators make the motion quasi-periodic and eventually leads to  chaotic motion (Fig \ref{fig:chaos}). In the next section we study the details of the time evolution through phase space sections and discuss the issue of non-integrability. It is to be noted that for very high value of one angular momentum (i.e $\dot \theta$) the space virtually shrinks in those direction and the motion again becomes apparently integrable. This is related to the fact that for very high energy motion one may neglect the details of the potential.

One interesting special case is $\alpha_1=\alpha_2$. Here we have a line of extrema along $\theta_1=\pm\theta_2$, although this line of extrema always has an unstable direction. Any motion which starts with $\theta_1=\pm \theta_2$ and $\dot \theta_1=\pm \dot \theta_2$ remains on the line and essentially becomes a motion with one parameter and is integrable. Also the potential has enhanced symmetry. Near the minimum of the potential there is an approximate rotational symmetry valid upto quartic order, which may lead to some interesting solutions.



\subsection{Poincar\`e sections and KAM theorem}

 In an integrable system the number of  conserved quantities equals  the number of degrees of freedom.
 A convenient way to understand those conserved charges is to  look at the phase space. Let us assume that we have an integrable systems with $N$ position variables $q_i$ with conjugate momenta $p_i$. The phase space is $2N$ dimensional. Integrability means that there are conserved charges $Q_i=f_i(p,q)$ which are constant of motion. One of them is energy. These charges define an $N$-dimensional surface in the phase space which is a topological torus (KAM tori). The $2N$ dimensional phase space is nicely foliated by these $N$ dimensional tori. In terms of action angle variables ($I_i,\theta_i$) these tori just become surfaces of constant action. With each torus there are associated $N$ frequencies $\omega_i(I_i)$ which are the frequencies of motion in one particular action direction.

What happens to those tori when the integrable hamiltonian is perturbed by a small non-integrable piece ? The KAM theorem states that most tori would be deformed but survive a small non-integrable deformation. However resonant tori which have rational ratio of frequencies, i.e. $m_i \omega_i=0$ with $m \in {\cal Q}$, will be destroyed. As the strength of the non-integrable interaction increases more tori gradually get destroyed. A nicely foliated picture of the phase space is no longer applicable and the time evolution may freely explore the whole phase space only constrained by energy. In such cases the motion becomes completely chaotic.

To numerically investigate this gradual disappearance of foliation we look at the Poincare sections of our system. Our phase space has four variables $\theta_1,\theta_2,p_{\theta_1},p_{\theta_2}$. If we fix energy we would be in a three dimensional subspace. Now if we start with some initial condition and time evolve, the motion would be confined to a two dimensional torus for the integrable case. This 2d torus intersects the $\sin(\theta_2)=0$ hyperplane at two circles. Taking gradual snapshots of the system as it crosses $\sin(\theta_2)=0$ and plotting the value of $(\sin(\theta_1),\cos(\theta_1)p_{\theta_1})$, we can reconstruct those circles. Furthermore varying the initial boundary condition, in particular keeping $(\theta_2(0)=\pi,p_{\theta_1}=0)$ and varying $\theta_1(0)$ to determine $p_{\theta_2}(0)$ from the energy constraint, we may hope to get a foliation structure typical for a integrable system.

\begin{figure}[htp]
\centering
\subfigure[KAM tori]{
\includegraphics[scale=0.45]{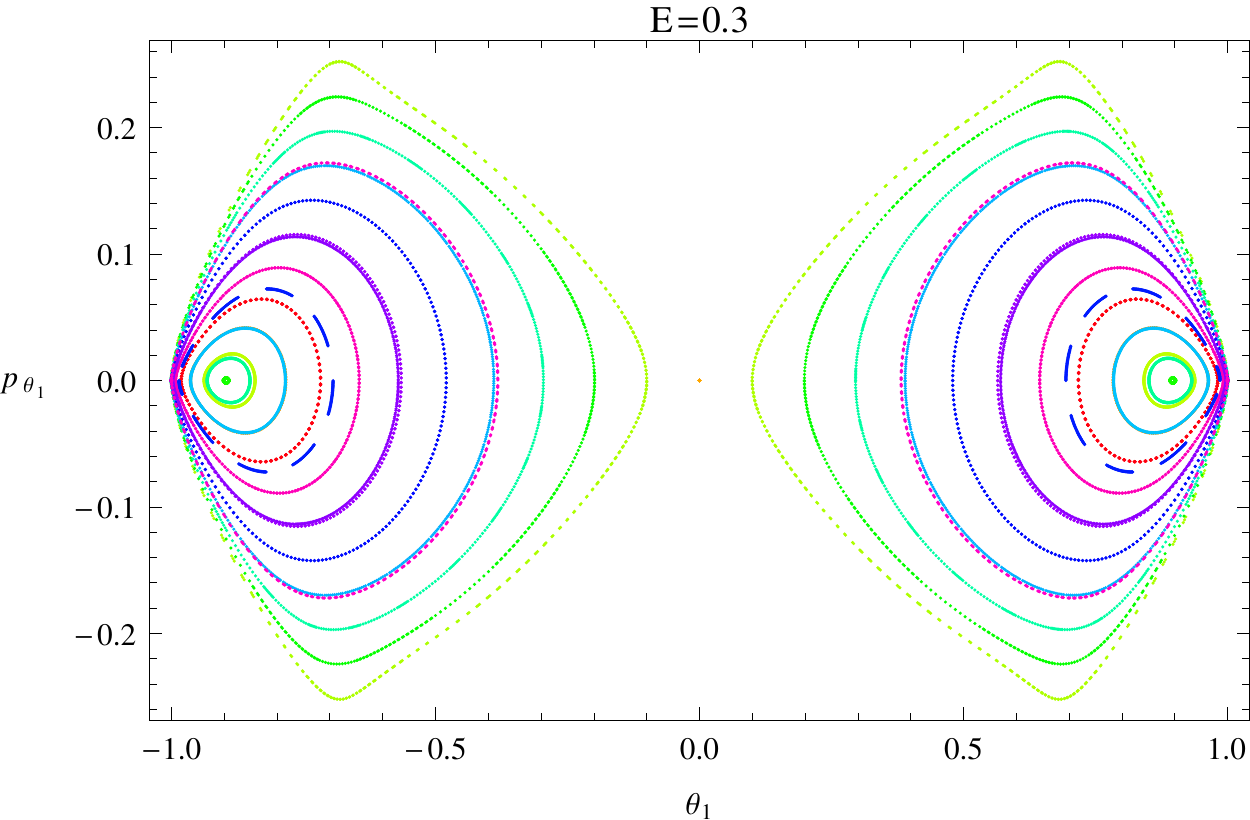}
\label{fig:psec1}
}
\subfigure[Visible chaos]{
\includegraphics[scale=0.45]{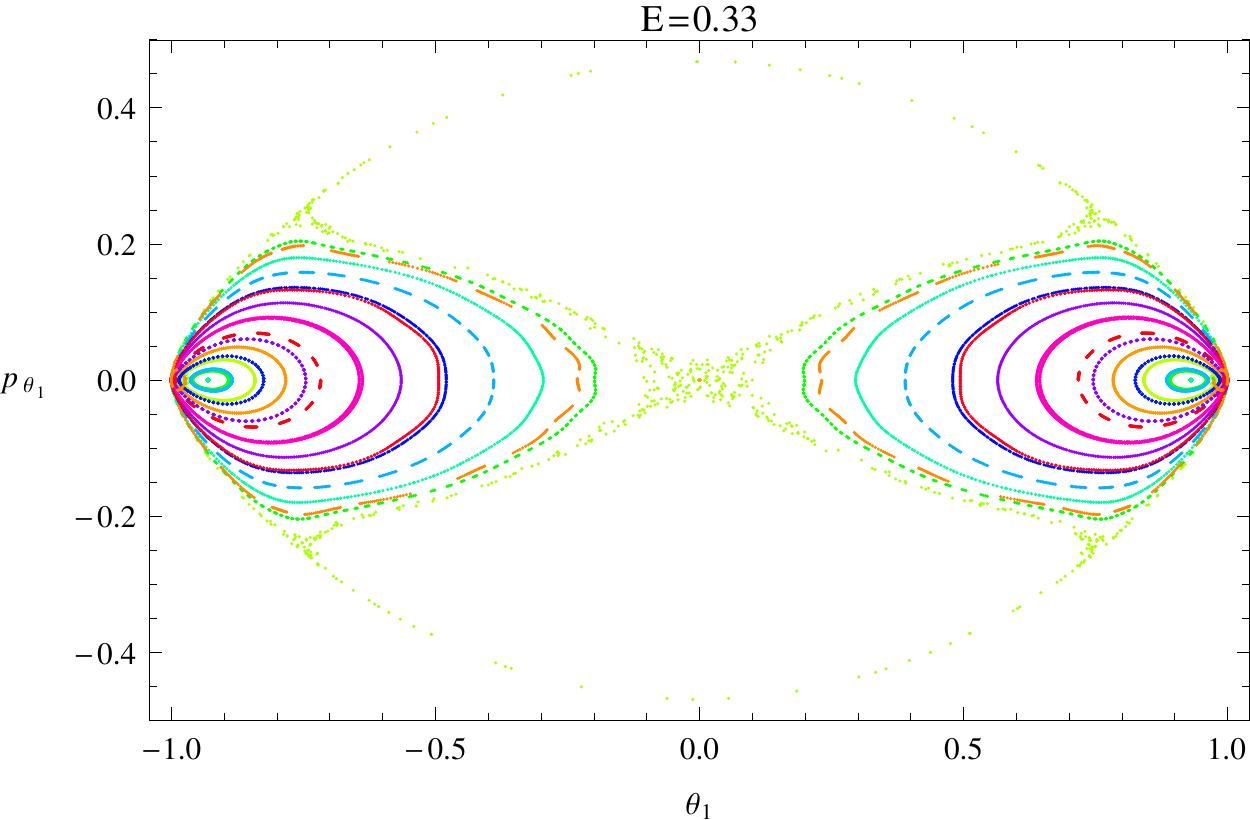}
\label{fig:psec2}
}
\subfigure[More into chaos]{
\includegraphics[scale=0.45]{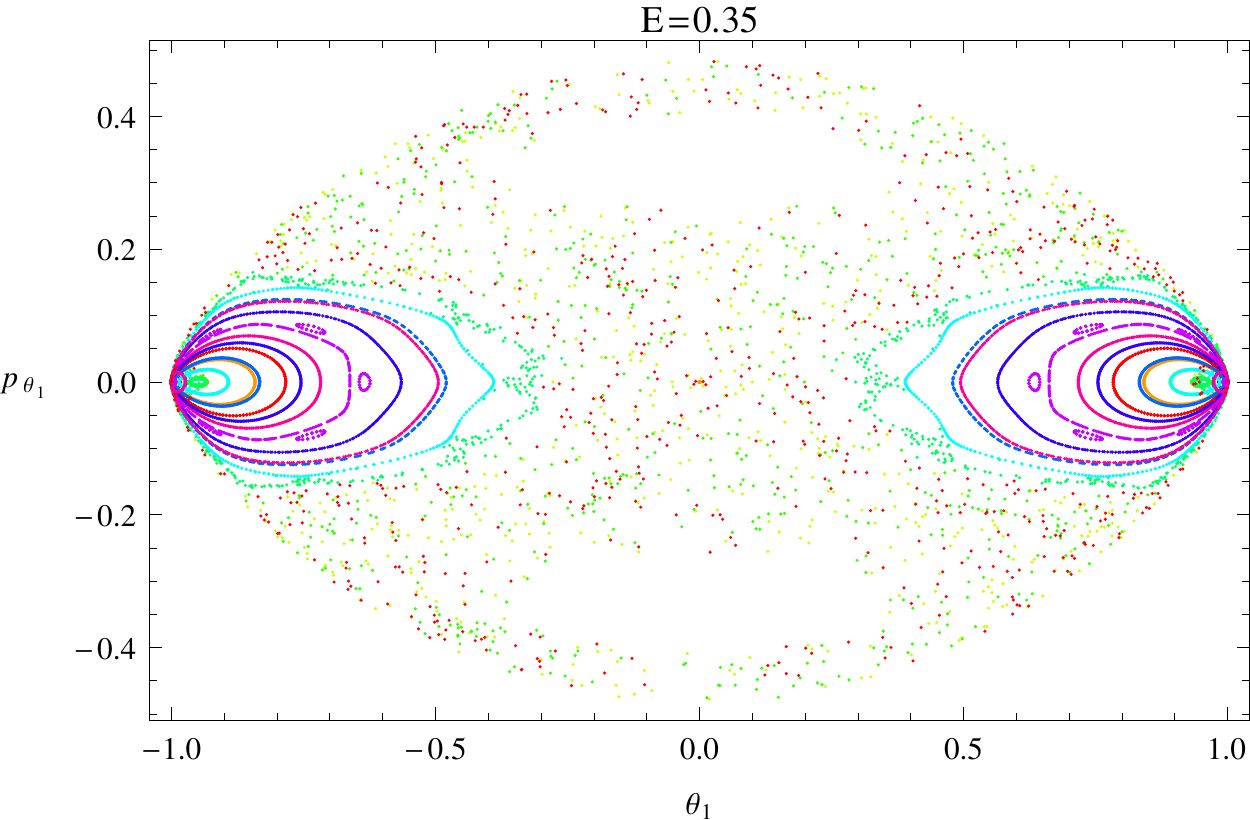}
\label{fig:psec3}
}
\subfigure[Even more into chaos]{
\includegraphics[scale=0.45]{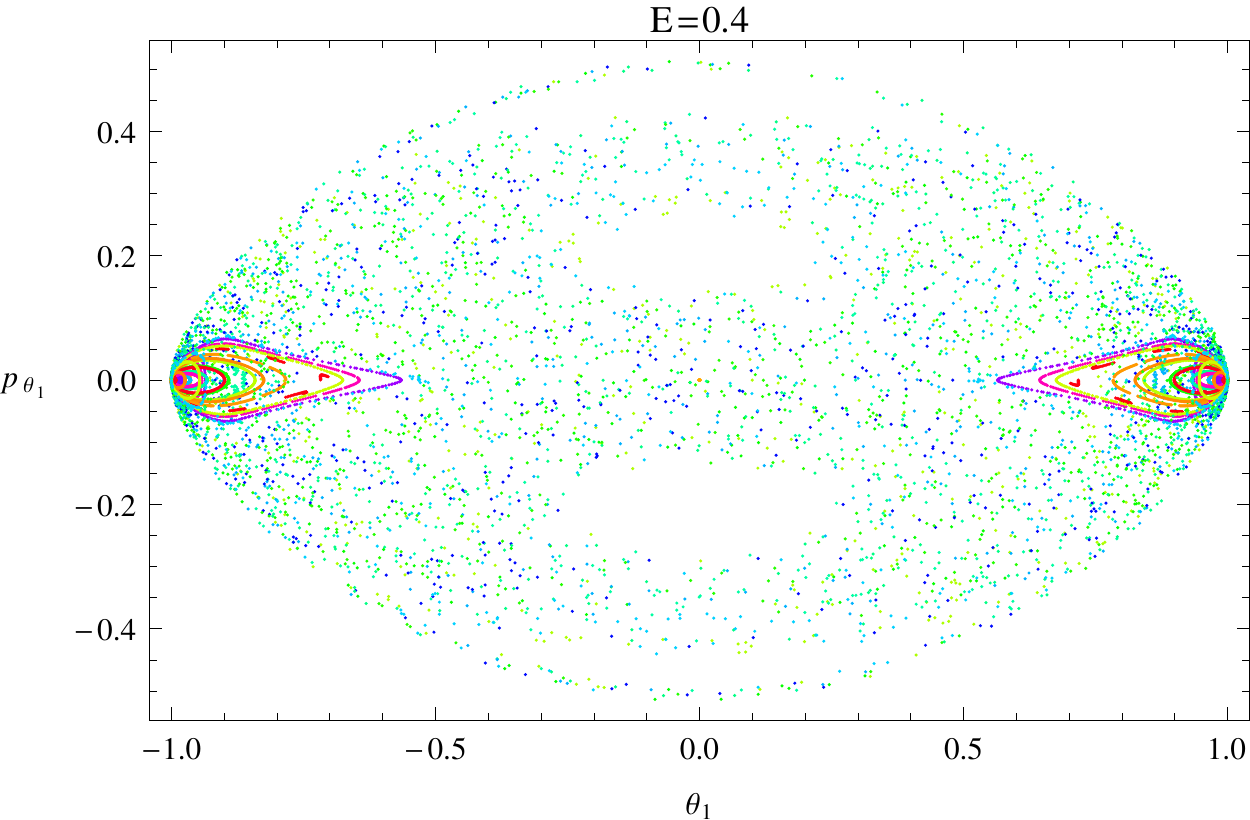}
\label{fig:psec4}
}
\subfigure[Completely chaotic]{
\includegraphics[scale=0.45]{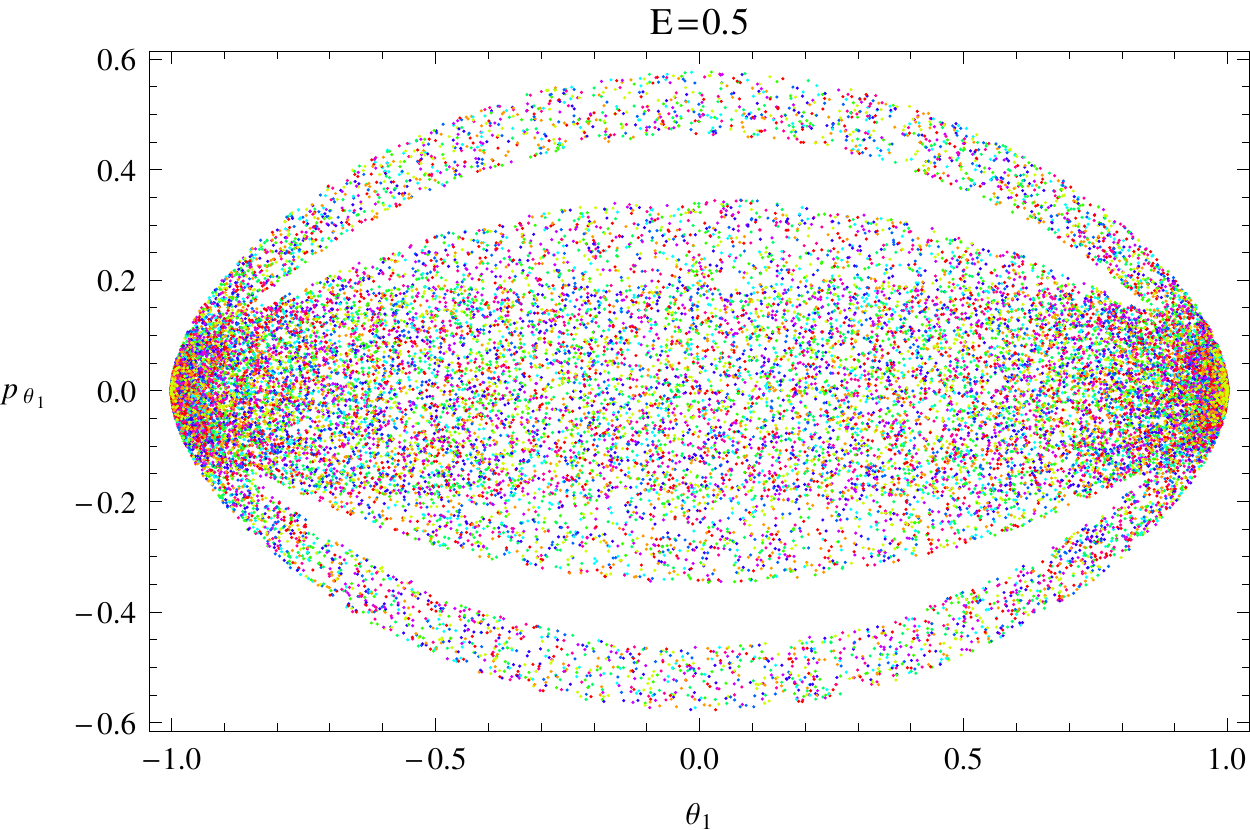}
\label{fig:psec5}
}
\subfigure[Formation of order]{
\includegraphics[scale=0.45]{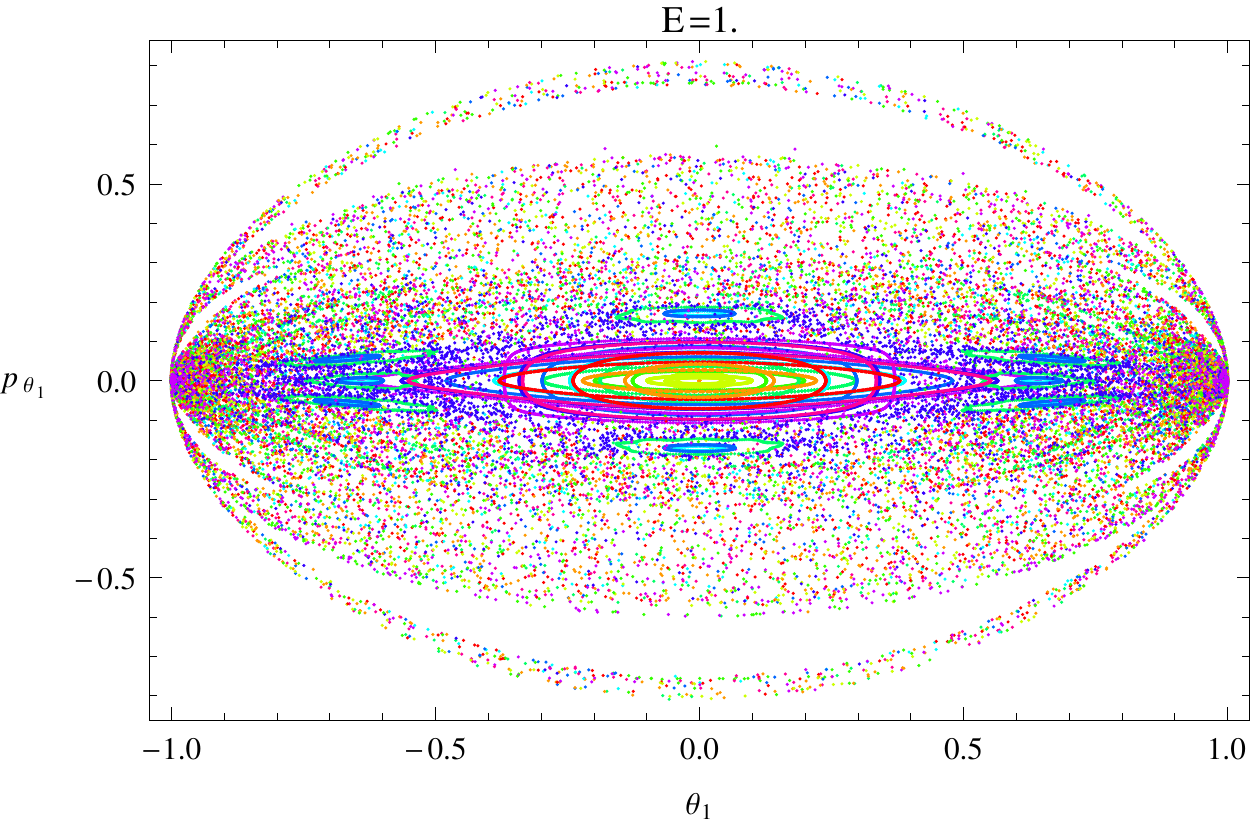}
\label{fig:psec6}
}
\subfigure[Ordered agian]{
\includegraphics[scale=0.45]{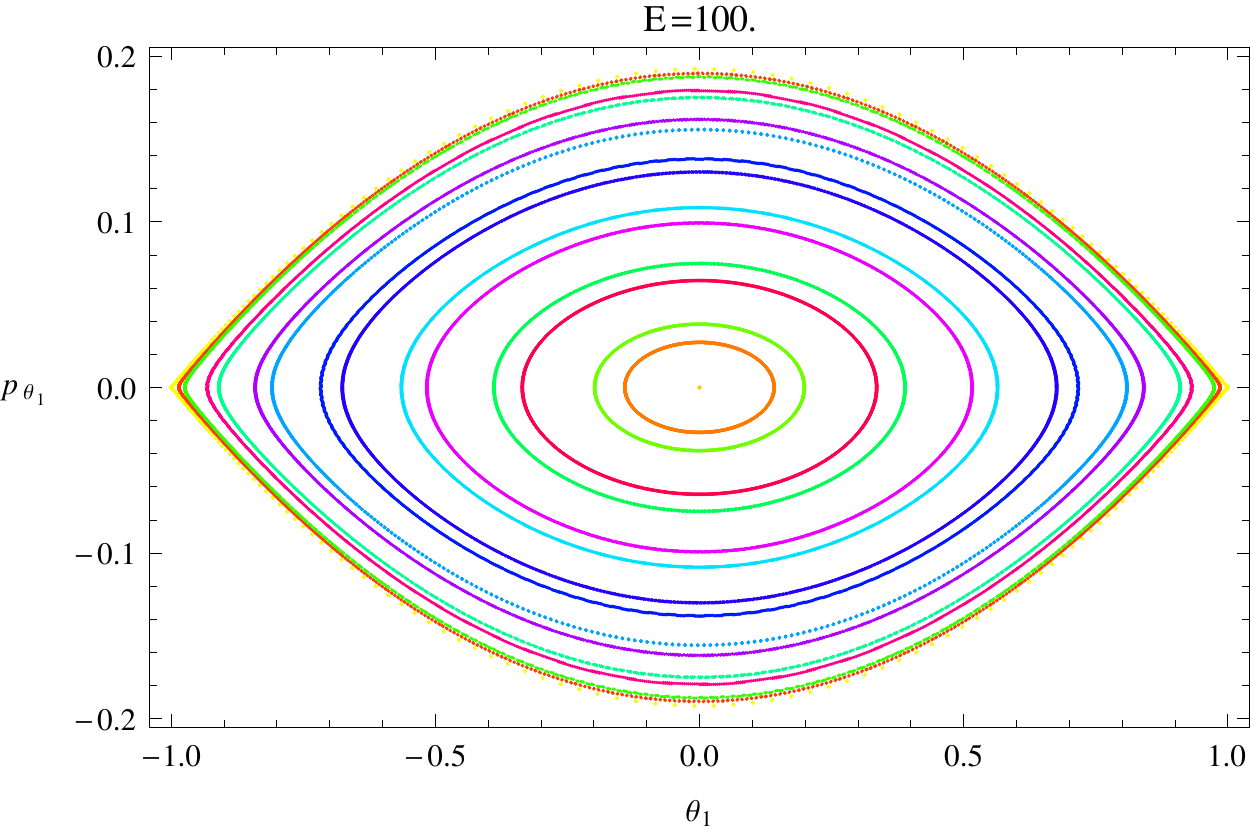}
\label{fig:psec7}
}\caption{Poincar\'e Sections.}
\end{figure}

Indeed we see that for smaller values the of energy, a distinct foliation structure exists in the phase space [Fig.\ref{fig:psec1}].  As we increase the energy some tori get gradually dissolved [Figs.\ref{fig:psec2}-\ref{fig:psec5}]. The tori which are destroyed sometimes broken down into smaller tori [Fig.\ref{fig:psec4}] eventually the tori disappear and become a collection of scattered points known as cantori. However the breadths of these cantori are restricted by the undissolved tori and other dynamical elements. Usually they do not span the whole phase space [Figs.\ref{fig:psec2}-\ref{fig:psec4}].  For sufficiently large values of energy there are no well defined tori [Fig.\ref{fig:psec5}]. In this case phase space trajectories are all jumbled up and trajectories with very different initial conditions come arbitrary close to each other [Fig.\ref{fig:psec5}]. Interestingly for even higher energy order seems to have again form in the system [Figs.\ref{fig:psec6}-\ref{fig:psec7}]. This is related to the fact that at very high energy we can neglect the potential. The mechanism is very similar to what happens in well known non-integrable periodic systems like double-pendulum models \cite{Ott,Hilborn}.

\subsection{Lyapunov exponent}
\begin{figure}
\centering
\subfigure{
\includegraphics[scale=.7]{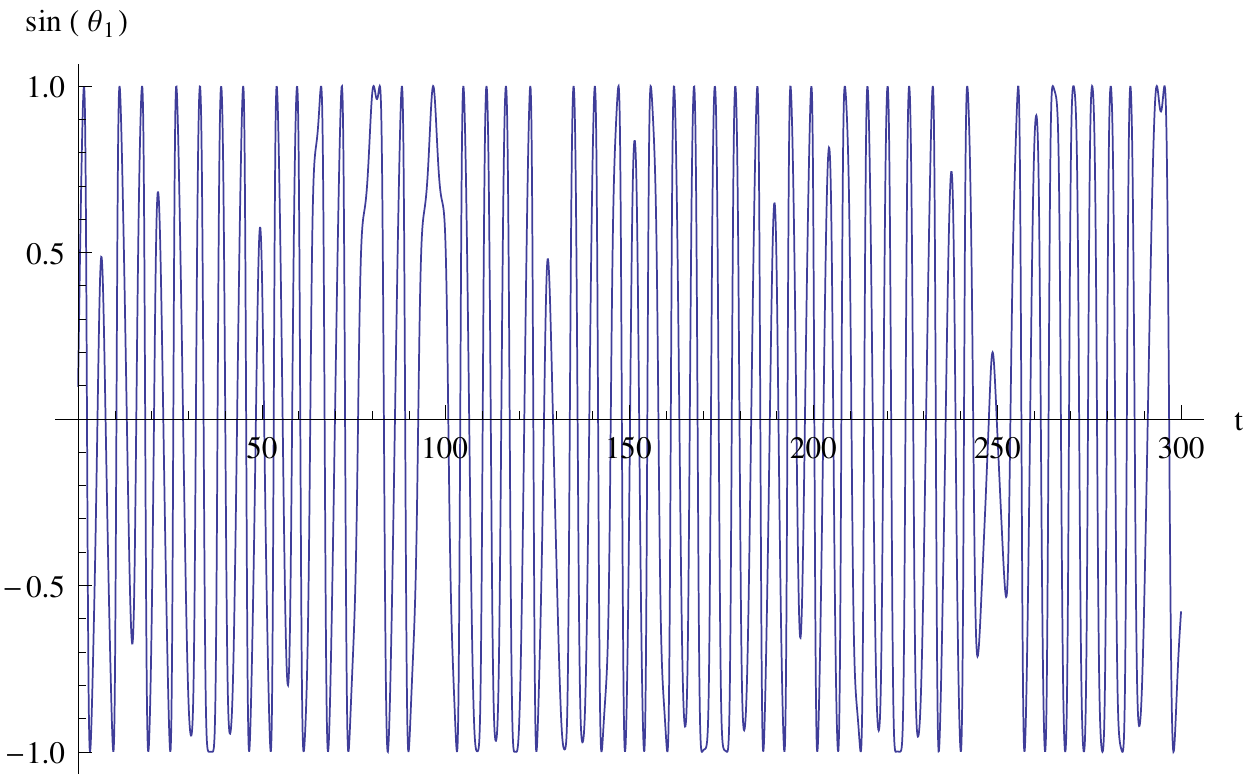}
}
\subfigure{
\includegraphics[scale=1.0]{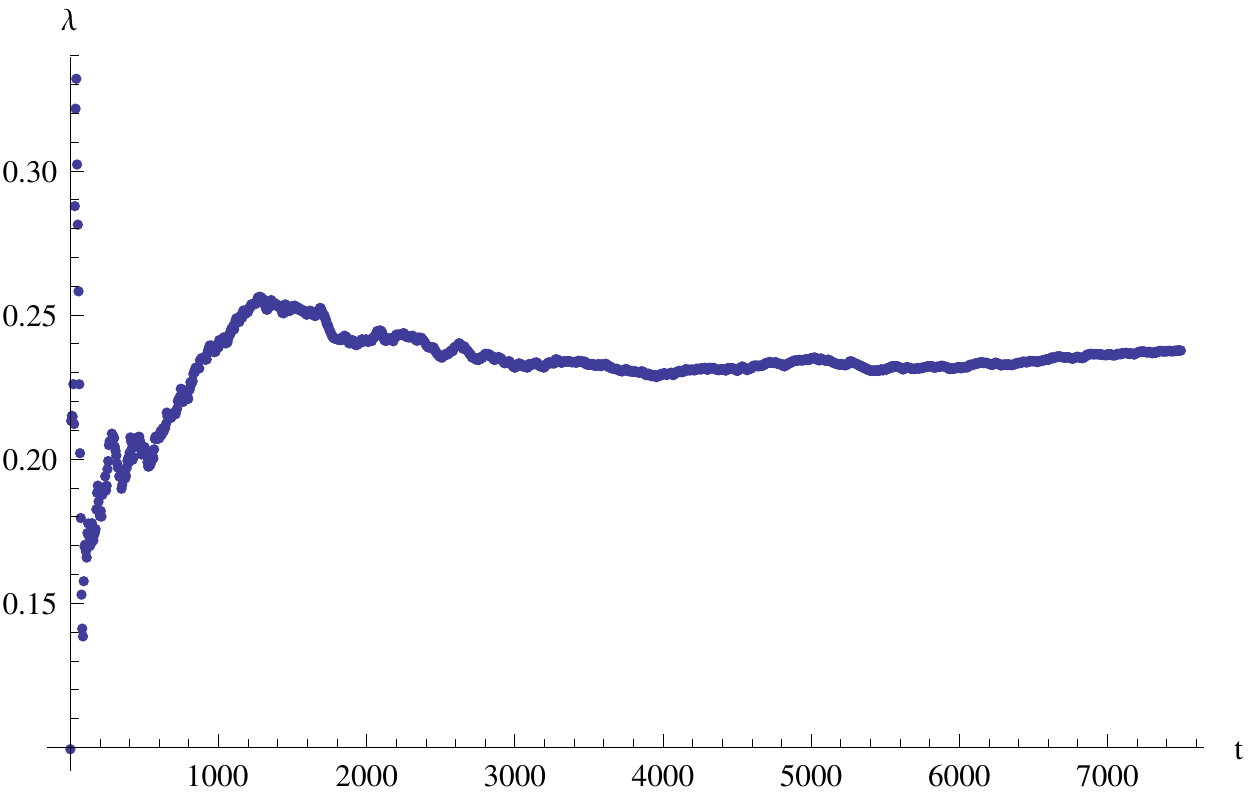}
}
\caption{Motion of the string and the corresponding Lyapunov indices for a chaotic motion. We have chosen the initial condition $\theta_1(0)=0.1,\theta_2=0.1,\dot \theta_1=0,\dot \theta_2=0.8$. We see a convergent Lyapunov index with $\lambda \approx 0.245$.}
\label{fig:liu}
\end{figure}

One of the trademark signatures of chaos is the sensitive dependence on initial conditions, which
means that for any point $X$ in the phase space, there is (at least) one point arbitrarily close to $X$ that diverges from $X$. The separation between the two is also a function of the initial location  and has the
form $\Delta X(X_0,\tau)$. The Lyapunov exponent is a quantity that characterizes the rate of separation of such infinitesimally close trajectories. Formally it is defined as,
\begin{align}
\lambda=\lim_{\tau\rightarrow \infty} \lim_{\Delta X_0 \rightarrow 0} \frac{1}{\tau} \ln \frac{\Delta X(X_0,\tau)}{\Delta X(X_0,0)}
\end{align}
In practice we use an algorithm by Sprott \cite{jcsprott,sprott1}, which calculates $\lambda$ over short intervals and then takes a time average. We should expect to observe that, as time $\tau$ is increased, $\lambda$ settles down to oscillate around a given value. For trajectories belonging  to the KAM tori, $\lambda$ is zero, whereas it is expected to be non-zero for a chaotic orbit. We have verified such expectations for our case. We calculate $\lambda$ with various initial conditions and parameters. For apparently chaotic orbits we observe a nicely convergent positive $\lambda$ (Fig. \ref{fig:liu}).

\section{ Comments on the dual field theory }\label{sec:fieldtheory}

The corresponding ${\cal N}=1$ superconformal gauge theory dual to $AdS_5\times T^{1,1}$ is known as the Klebanov-Witten theory and was originally described in \cite{Klebanov:1998hh}. The theory has  flavor symmetry $SU(2) \times  SU(2)$.  The elementary degrees of freedom are denoted by the fields $A$ and $B$, each a doublet of the factor groups $SU(2)\times SU(2)$ and with
conformal anomalous dimension $\Delta_{A,B} = 3/4$. The gauge group is $SU(N) \times SU(N)$
and the two chiral multiplets $A$ and $B$ are respectively in the $(N,\bar{N})$ and $(\bar{N},N)$. The superpotential is 
\be
W={\lambda\over 2}\ep^{ij}\ep^{kl}Tr[A_iB_kA_jB_l], 
\ee
with $i=1,2$.

The chiral operators analogue of the $(X,Y, Z)$ operators in ${\cal N}=4$ SYM are in this case given by $Tr(AB)^k$ with R-charge $k$ and in the
$( \frac{k}{2}, \frac{k}{2})$ representation of the flavor group $SU(2) \times  SU(2)$.

A discussion of the field content in relation to classical variables can be found, for example, in \cite{PandoZayas:2002rx} whose notation we follow. As our starting point we can use a BMN-like operator of the form

\be
Tr[(A_+B_+)^J]
\ee
This is an operator describing a spinning string in $T^{1,1}$. In our case $J$ corresponds to the angular momentum along the $\psi$ direction. This is a supersymmetric operator as it follows from the bounds of the supersymmetry algebra presented in \cite{Ceresole:2000rq}. Roughly speaking we associate angular momentum along $\phi_1$ and $\phi_2$ in $T^{1,1}$ with the insertion of opertors of the form $A$ and $B$. Let us schematically represent these operators as $|k_1,\k_2>$. These are, of course, not yet the kind of operators we are interested as we look for operators describing winding rather than momentum. Thus we Fourier transform this operators along the same lines as for the magnon operatorn. Namely we consider operators of the form $|\alpha_1,\alpha_2>=\sum e^{ik_1 \alpha_1 +i k_2 \alpha_2}|k_1, k_2>$. To summarize, we take  BMN-like operators and insert impurities that carry quantum numbers associated with angular momentum in the $\phi_1$ and $\phi_2$ directions and then perform a Fourier transform.

\section{ Conclusions }\label{sec:conclusions}

In this paper we have shown that certain classical configurations corresponding to a string winding along two of the angles of $T^{1,1}$ exhibits chaotic motion for some range of initial conditions. This is enough to claim that the classical string in $AdS_5\times T^{1,1}$ can not be integrable and therefore we settle the important question of the existence of integrability beyond the known example of string theory on $AdS_5\times S^5$. We have thus showed that integrability seems to be a very peculiar property of $S^5$, changing that manifold by what is considered the simplest next example (modulo orbifolds of $S^5$) leads us into the realm of chaotic solutions. Given the simplicity of our dynamical system (two coupled gravitational pendula) we are very confident in the robustness of the numerical result which precisely matches what is expected from other text-book type of examples. Our example is actually very similar to the H\'enon-Heiles system and we exploited this similarity to make our case for chaotic behavior completely convincing following the standard literature on dynamical systems \cite{Ott,Hilborn,jcsprott}.

As we saw in section \ref{sec:analysis} there are some islands, although of measure zero, of the phase space of the motion of the string in $AdS_5\times T^{1,1}$ where the system is integrable. For example, allowing the string to wrap only one direction in $T^{1,1}$ yields an integrable system; it is only when we consider a configuration wrapping {\it both} directions $\phi_1$ and $\phi_2$ that we find chaotic behavior.

It is worth remarking that the classical particle in $AdS_5\times T^{1,1}$ is integrable. The simplest way to see this is directly from the separability of the Laplacian. Of course, this is precisely the regime necessary for the study of supergravity modes. The implications of this situation - integrability of particles and chaos for strings -  are translated in the dual field theory as corresponding to properties of operators dual to stringy states. Generically these are operators with very large quantum numbers and are not protected by symmetries. Recall, for example,  that in the case of the BMN operators the classical string configuration consists of string shrunk to a point orbiting along a large circle in $T^{1,1}$. The ground state operator is protected and the main point of BMN is that a subset of perturbations or insertions of operators remains semi-protected. What we anticipated due to the onset of chaotic behavior is that there are many operators for which insertions can not form a closed set.

One interesting implication of our calculation is that it is easily generalizable to other Sasaki-Einstein manifolds such as $Y^{p,q}$ and $L^{p,q,r}$. As we have seen in section \ref{sec:analysis} a key property of the manifold that translated into the coupling of the corresponding oscillators was the existence of a nontrivial fibration. More precisely, the interaction contribution to the potential term which takes the form $\cos\theta_1\cos\theta_2$ in equation (\ref{eq:potential}) arises precisely because the space is a nontrivial $U(1)$ fibration over $S^2\times S^2$. The corresponding effect is present in the generic Sasaki-Einstein 5-manifold which can be written as a nontrivial $U(1)$ fibration over a K\"ahler-Einstein base.


\section*{Acknowledgments}
We thank J. T. Liu, C. Terrero-Escalante and  D. Reichmann for comments. P.B. thanks Archisman Ghosh, Diptarka Das, Sumit Das, Al Shpere and especially Spenta Wadia for discussion. P.B. thanks TIFR, Mumbai and MCTP in Ann Arbor for hospitality during the initial stage of this work. This work is  partially supported by Department of Energy under grant DE-FG02-95ER40899 to the University of Michigan. P.B. is supported by National Science Foundation grant NSF-PHY-0855614.

\end{document}